# Data Privacy Preservation on the Internet of Things

Jaydip Sen[*] & Subhasis Dasgupta
Department of Data Science, Praxis Business School, Kolkata, INDIA
Corresponding author's email: [*]jaydip.sen@acm.org

## 1. Introduction

Recent developments in hardware and information technology have enabled the emergence of billions of connected, intelligent devices around the world exchanging information with minimal human involvement. This paradigm, known as the Internet of Things (IoT) is progressing quickly with an estimated 27 billion devices by 2025 (almost four devices per person) [1-2]. These smart devices help improve our quality of life, with wearables to monitor health, vehicles that interact with traffic centers and other vehicles to ensure safety, and various home appliances offering comfort. This increase in the number of IoT devices and successful IoT services has generated a tremendous amount of data. The International Data Corporation report estimates that by 2025 this data will grow from 4 to 140 zettabytes [3].

However, this humongous volume of data poses growing concerns for user privacy. Gartner predicts approximately 15 billion connected devices will be linked to computing networks by 2022 [4]. Not only could these gadgets be vulnerable but also the massive amounts of unsecured data stored online create a liability. Users having difficulty controlling the data from their devices has highlighted privacy as a major issue. To guarantee high levels of user data protection, IoT systems must adhere to regulations such as the European Union's *general data protection regulation* (GDRP) of 2018 [5]. These regulation policies focus on giving users control over what is collected, when, and for what purpose. By 2023, the regulators will demand organizations protect consumer privacy rights for more than 5 billion citizens and comply with more than 70% of the GDPR requirements [5].

Traditional privacy protection schemes are insufficient for IoT applications which necessitate new techniques such as distributed cybersecurity controls, models, and decisions that take into account vulnerabilities in system development platforms as well as malicious users and attack surfaces. Machine learning techniques can provide improved detection of novel cyberattacks when dealing with large volumes of data in IoT systems. Furthermore, they can enhance how sensitive data are shared between components to keep them secure. Machine learning-based schemes thus improve the operations related to privacy protection and more effectively comply with the regulations. In this chapter, a survey is presented on the currently existing machine learning-based approaches for the privacy preservation of data in the IoT.

The rest of the chapter is organized as follows. Section 2 identifies some of the existing surveys on the privacy preservation of data in the IoT. Section 3 discusses current privacy schemes for IoT based on a centralized architecture. Section 4 highlights the existing schemes working on the principles of distributed learning. In Section 5, some well-known privacy schemes on distributed encryption mechanisms are discussed. The concept of differential privacy and some schemes working on this principle are presented in Section 6. Finally, the chapter is concluded in Section 7 highlighting some emerging trends in the field of privacy in the IoT.

## 2. Some Existing Survey Works on Privacy Issues in IoT

In the literature, several studies have reviewed privacy issues in IoT environments, focusing mostly on threats and attacks on such systems. A comprehensive survey is carried out on various threat models and the classification of various attack types in the context of IoT [6]. The study found that the training dataset used for building the machine learning model for designing the privacy protection system is the most vulnerable asset to a possible attack. Other sensitive assets are the model, the parameters and hyper-parameters involved, and the model architecture. On the other hand, the sensitive actors are the owners of data, the owners of the model, and the users of the model. Another important observation of this study is that among the machine learning models, the ordinary least square regression model, decision tree, and support vector machine model are the most vulnerable ones. Another recently published paper presented a comprehensive survey on various machine learning and deep learning-based approaches used for protecting the privacy of user data in the IoT [7].

Many surveys focus on reviewing the mechanisms and models for preserving the privacy of data. Various issues that are considered include differential privacy, homomorphic encryption, and learning architectures and models. In one study, the threats and vulnerabilities of privacy protection systems in IoT are classified into four groups (i) attacks on authentication, (ii) attacks on the components of edge computing, (iii) attacks on the anatomization and perturbation schemes, and (iv) attacks on data summarization [8]. In another survey work, the existing privacy protection systems with centralized architectures and machine learning approaches are analyzed by categorizing the data generated at different layers [9]. Kounoudes & Kapitsaki [10] analyzed several privacy preservation solutions to determine basic characteristics. The authors proposed a mix of machine learning techniques for providing user protection, along with the policy languages to set user privacy preferences, and negotiation techniques that improve services while preserving user rights. Zhu et al.'s survey work included several approaches including differential privacy, secure multi-party computing, and homomorphic encryption for training models [11]. The authors classified the models based on collaborative or aggregated scenarios for the protection of user identity or information. Ouadrhiri et al. analyzed the current methods within federated learning environments classifying them into three distinct groups: (i) *k*-anonymity, (ii) *l*-diversity, and (iii) *t*-closeness to protect datasets [12]. It is also observed by the authors that differential privacy-based technologies are mostly used for training the privacy models. This approach, however, suffers from a high computational complexity for the encryption and decryption operations.

## 3. Centralized Architecture-Based Encryption Schemes

The data privacy mechanisms and systems under this category make use of encryption techniques such as homomorphic encryption, attribute access control, multi-party computation, and lightweight cryptography. These approaches are usually resource hungry and involve high computational resources and large memory spaces. Homomorphic encryption systems provide a very high level of privacy even when it is deployed in third-party computations. Researchers designed several variants of homomorphic encryption systems such as partially homomorphic encryption, and somewhat homomorphic encryption [13-15]. While somewhat homomorphic encryption systems minimize communication overhead by using a smaller key size, partially homomorphic encryption systems are suitable for lightweight protocols of IoT since they yield shorter ciphertexts.

In building privacy models, the modelers encounter a difficult challenge. While data owners do not what to expose their sensitive information to untrusted and potentially malicious models, the model owners prefer not to share information about their models as they are valuable assets. As such, classification protocols utilize machine learning classifiers over encrypted data to protect privacy on both sides. Bost et al., De Cock et al., Rahulamathavan et al., Wang et al., Zhu et al., and Jiang et al., all proposed several protocols for privacy-preserving classification using different datasets and models including hyperplane decision, naive Bayes, decision trees, support vector machines, multi-layer extreme learning machine amongst others. These models have yielded an accuracy of results varying between 86-98% [16-21]. These efforts also result in reduced training and execution times compared to traditional deep learning models like convolutional neural networks.

## 4. Privacy Schemes Using Distributed Learning

Of late, privacy protection of data using distributed machine learning [22-23] has gained considerable popularity in the context of the IoT. Distributed machine learning allows the learning models to be generated at each participant device, while the central server acting as the coordinator creates a global model and distributes the knowledge to the participating nodes. Shokri and Shmatikov proposed a collaborative computing system that works on deep learning to protect the sensitive data of a user while utilizing the information content of non-sensitive data of other users in the system [22]. The deep learning algorithms use the stochastic gradient descent algorithm because of the parallelization and asynchronous execution capability of the latter. The privacy model is found to yield a very high accuracy on the test dataset. A distributed learning-based mechanism for data privacy preservation on IoT devices is proposed by Servia-Rodriguez et al. [23]. The scheme does not involve any data communication to the cloud environment. The system works in two phases. In the first step, the model is trained on data voluntarily shared by some users and possibly not containing any privacy-sensitive information. Once the model is trained, no further user data are shared. The model, tested on a public dataset, yields high accuracy. This scheme assumes user data privacy preservation since the original data never leaves their device. However, this is incorrect as the distributed machine learning models are vulnerable to privacy inference attacks that attempt to access privacy-sensitive data or model inversion attacks recovering original data [24-25]. This enforces protection techniques such as encryption or differential privacy into distributed learning systems.

## 5. Distributed Learning and Encryption in Privacy Preservation

To boost data privacy in IoT applications, encryption techniques are integrated into distributed machine learning. The most commonly used encryption method used is homomorphic encryption in which the user data is encrypted before being sent to the computing nodes. A privacy protection system has been proposed based on the joint operation of a multi-layer perceptron and a convolutional neural network model [26]. The model has been tested on the *modified national institute of standards and technology* (MNIST) and *street view house number* (SVHN) datasets [42]. A secure information system for healthcare applications in the IoT environment has been proposed [27]. The proposed model uses Pallier additive homomorphic encryption [28]. Another privacy system based on the Pallier system has been presented that works on blockchain technology [29]. The authors tested the system on two datasets of the University of California, Irvine (UCI) data repository [30-31]. Homomorphic encryption systems offer increased privacy compared to differential privacy-based ones. However, fully homomorphic encryption can be costly in terms of computation overload, while partial homomorphic encryption can only be used for carrying out single operations. Moreover, partial homomorphic encryption methods require trusted third parties in place, or they work on simpler models approximating complex equations using single mathematical operations. A mechanism is proposed for protecting the privacy of data for the Industrial Internet of Things (IIoT) built on the principles of distributed learning [32]. The scheme works on a variational autoencoder model trained using homomorphic encryption. The accuracy of the model is found to be high, while its execution time is low. A hybrid framework for privacy protection is proposed by Osia et al. [33]. The scheme utilizes Siamese architecture and can perform efficient privacy-preserving analytics splitting a neural network IoT devices and cloud [34]. The feature extraction is done at the device, while the classification is carried out in the cloud. The scheme uses a convolutional neural network model evaluated with gender classification datasets *Internet movie database* (IMDB-Wiki) [35] and *labeled faces in the wild* (LFW) [36], achieving an accuracy of 94% & 93%, respectively. A data privacy-preserving scheme known named CP-ABPRE is presented by Zhou et al. that works on a policy-based encryption approach [37]. The scheme is found to be robust against privacy attacks and has a low computational overhead required for its encryption and decryption processes.

## 6. Privacy Schemes Using Distributed Learning and Differential Privacy

In the differential privacy approach, the privacy of data is protected through the addition of some random perturbations into the original data. In other words, a perturbation in the data is done with a predetermined measure of the error caused by modifications to the data [38]. There are several well-known techniques of perturbations such as swapping, randomized response, micro-aggregation, additive perturbation, and condensation. However, perturbations reduce the quality of the data for analysis as the original data are modified. Privacy models work on a trade-off between the utility of data and its associated privacy level. In the privacy-utility tradeoff, several algorithms and approaches exist in the literature. In the context of differential privacy, Abadi et al. presented a scheme involving the training of a neural network with differential privacy for preventing the disclosure of sensitive information [39]. The scheme is proved to be highly effective in preserving the privacy of sensitive data as observed from its performance on the test dataset. Another scheme for privacy preservation of sensitive data is proposed in which a subset of parameters is shared and obfuscated using differential privacy as the training of the deep learning structures is carried out locally [40]. While the differential privacy-based schemes do not need high computational resources, they may be inaccurate since perturbations can reduce training quality. Moreover, these schemes cannot fully protect data privacy (i.e., there is always a tradeoff made between the accuracy-privacy of the model). Wang et al. [41] enhanced the performance of the distributed machine learning system with differential privacy in an IoT environment via their Arden framework [41]. The scheme proposed by the authors involves protecting sensitive information using nullification or noise addition [42]. The model is tested on the MNIST/SVHN datasets and has been found to yield high accuracy while considerably reducing resource consumption [42]. The scheme proposed by Zhang et al. focused on distributed sensing systems where an obfuscate function was used to preserve training data privacy when shared with third parties [43].

Lyu et al. proposed a privacy mechanism that used the random projection method for perturbing the original data and embedding fog computing into deep learning [44]. This scheme is able to reduce communication overhead and computation load. The novel method of privacy protection known as the fog-embedded privacy-preserving deep learning framework can preserve the privacy of data using a robust defense method. First, a random perturbation is used to perturb yet preserve the statistical characteristics of the original data. Then, differentially private stochastic gradient descent is used to train the fog-level models with a multi-layer perceptron model. The multi-layer perceptron model consists of two hidden layers equipped with the *rectified linear unit* (ReLU) activation function. The accuracy yielded by the scheme on the test data is found to be quite acceptable, although it is slightly lower in comparison to those of models with centralized architecture. However, the communication and computation overheads are significantly reduced.

Some privacy preservation schemes utilize Gaussian projections to efficiently implement collaborative learning environments [45]. In these schemes, the resource-constrained IoT devices participate collaboratively and randomly apply multiplicative Gaussian projections on the records of the training data. This process obfuscates the privacy-sensitive input data. The coordinator node applies a deep learning-based model to learn from the complex patterns of the obfuscated data supplied by the Gaussian random projections. The performance results of the scheme demonstrated its efficiency and effectiveness in data privacy protection.

Among other approaches, obfuscation-based methods are also used in distributed machine learning to control the overhead of computation involved in the encryption procedures in massively large-sized data. A scheme proposed by Alguliyev et al. protects big data in the context of IoT [46]. The mechanism involves the transformation of sensitive data into data that can be publicly shared. The proposed method works in two phases. In the first phase, data is transformed by passing it through a denoising type autoencoder. The parameter for designating the sparsity parameter of the autoencoder is specified for minimizing the loss in the autoencoder objective function during the data compression process. In the second phase, the transformed data from the output of the denoising autoencoder is classified using a convolutional neural network model. The proposed scheme was tested on several disease datasets and was found to be highly accurate in its prediction. Du et al. proposed a novel privacy-preserving scheme for big data in IoT deployed in edge computing applications [47]. The mechanism is based on a differential privacy approach built on machine learning models which can improve query accuracy while minimizing the exposure of sensitive data to the public. The working mechanism involves two steps. In the first step, a Laplacian noise is added to the output data to carry out perturbation, while in the second step, random noise is added to the objective function that reduces the disturbance to the objective values. The data perturbation is carried out before transferring the data to the edge nodes. The model is tested on four diverse datasets and is found to be highly accurate in its performance. The machine learning models used in the scheme are stochastic gradient descent and generative adversarial networks.

Speech recognition systems, which are commonly found in IoT services, are susceptible to breaching user privacy as voice information is generally transmitted as plaintext and sometimes used for authentication purposes. To address this issue, Rouhani et al. proposed a scheme called *deepsecure* [48]. The working principle of the scheme is based on the garbled circuit protocol of Yao [49], and it executes much faster in comparison to the homomorphic encryption-based schemes. However, the proposition suffers from issues related to reusability and difficulty in implementation [50]. Differential privacy has been utilized in a work by adding perturbations to user data [39]. However, the proposed scheme has a lower level of accuracy. Ma et al. [51] have thereby improved upon this by proposing a secret-sharing-based method that improves accuracy and reduces the computation and communication overhead for both linear and nonlinear operations using a long-and-short-term memory network model with interactive protocols for each gate. The proposed scheme was tested on a private dataset yielding a very high accuracy. Although privacy preservation approaches based on obfuscation methods, in most cases, overcome the shortcomings of distributed machine learning and encryption-based distributed machine learning methods, these schemes are found to be vulnerable to some attacks [52-54].

## 7. Conclusion

This introductory chapter has presented a brief survey of some of the existing data privacy-preservation schemes proposed by researchers in the field of the Internet of Things. However, the design of privacy protection schemes in resource-constrained devices is still in its early stages. Reducing the latency and throughput of neural network training on encrypted data for privacy protection is a big challenge. Most of the existing schemes deploy their deep learning tasks to some external resources with adequate computing resources and storage spaces while keeping user data protected, making the schemes computationally efficient. New approaches should explore alternatives, such as quantum computing techniques, for designing more efficient and precise systems. In terms of future possibilities, parallel learning and cost optimization are being pursued, like network pruning and how different malicious activities interact. Effective standardization efforts should also be made by the relevant standard bodies for all privacy protection schemes [55]. Finally, evaluating and assessing privacy solutions in real-world scenarios is tough, especially when considering the balance between IoT quality-of-service and privacy protection.